\documentclass[twocolumn,superscriptaddress,secnumarabic,amssymb, nobibnotes, aps, prb]{revtex4-2}

\usepackage{amsmath}    % need for subequations
\usepackage{graphicx}    % need for figures
\usepackage{verbatim}    % useful for program listings
\usepackage{color}           % use if color is used in text
\usepackage{subfigure}   % use for side-by-side figures
\usepackage{hyperref}    % use for hypertext links, including those to external documents and URLs
\usepackage{braket}
\usepackage{physics}
\usepackage{xfrac}
\usepackage{xspace}
\usepackage{hyperref}
\newcommand{\amt}{$\alpha$-MnTe\xspace}

\newcommand{\mf}{MnF$_2$\xspace}
\newcommand{\nf}{NiF$_2$\xspace}
\newcommand{\bk}{\mathbf{k}}
\newcommand{\hbk}{\hat{\mathbf{k}}}

\newcommand{\bs}{\mathbf{s}}

\newcommand{\bS}{\mathbf{S}}
\newcommand{\bh}{\mathbf{h}}
\newcommand{\bbm}{\mathbf{m}}

\newcommand{\bL}{\mathbf{L}}
\newcommand{\bB}{\mathbf{B}}
\newcommand{\hbL}{\hat{\mathbf{L}}}
\newcommand{\na}{[110]\xspace}

\newcommand{\nc}{[001]\xspace}
\newcommand{\oT}{\operatorname{T}}
\newcommand{\bmD}{\boldsymbol{\mathcal{D}}}
\newcommand{\mD}{\mathcal{D}}

\def\bL{\mathbf{L}}

\def\hT{\hat{T}}

\begin{document}
%\title{X-ray Magnetic Circular and Linear Dichroism in Altermagnetic \mf}
\title{On the N\'eel Vector Dependence of X-ray Magnetic Circular Dichroism in Altermagnets}

\author{J.~Kune\v{s}}
%\affiliation{Institute for Solid State Physics, TU Wien, 1040 Vienna, Austria}
\affiliation{Department of Condensed Matter Physics, Faculty of
  Science, Masaryk University, Kotl\'a\v{r}sk\'a 2, 611 37 Brno,
  Czechia}
  \email{kunes@physics.muni.cz}

\begin{abstract}
Dependence of x-ray magnetic circular dichroism on the experimental geometry is described by a frequency-dependent Hall vector.
Using group theory, we derive a general relationship between the Hall vector and the orientation of the N\'eel vector $\bL$ in altermagnets within the free valence spin (FVS) approximation, where the spin-orbit coupling of the valence electrons and their exchange interaction with the core electrons are neglected.
For a given spin point group, the full $\bL$-dependence of the Hall vector can be expressed in terms of several irreducible spectral functions.
This derivation generalizes earlier results for the special cases of MnTe and MnF$_2$.
Depending on the system symmetry, XMCD in the FVS approximation may be present, emerge only when the neglected terms are included, or be completely forbidden.
\end{abstract}

\maketitle

%\section{Introduction}
\section{Introduction}

Altermagnetism~\cite{Smejkal22a, Smejkal22} has emerged as a rapidly growing area of solid-state physics. In addition to their characteristic spin-polarized electronic bands~\cite{Ahn19,Naka2019,Hayami19,Smejkal22a,Smejkal20,Yuan20,Yuan21,Hayami20,Smejkal22,Mazin21,Liu22,Yang2024}, altermagnets—previously classified as antiferromagnets—exhibit a range of distinguishing properties with promising applications. These include the anomalous Hall effect (AHE)\cite{Smejkal22b,Smejkal20,Samanta20,Naka20,Hayami21,Mazin21,Gonzalez2023,Naka22}, linear magneto-optical effects\cite{Naka20,Hariki2024a,Hariki2024b,Sasabe23}, and other phenomena~\cite{Watanabe2024} characterized by an odd dependence on the Néel vector.

In physics, rigorous classification typically relies on symmetry. For altermagnets, classification is achieved using non-relativistic spin groups~\cite{Smejkal22}, which describe systems in which orbital and spin degrees of freedom are decoupled. This framework—based on an approximate high-symmetry (non-relativistic) Hamiltonian perturbed by a symmetry-breaking term such as spin-orbit coupling (SOC)—is widely used in solid-state models, for example in spin-isotropic Heisenberg-type systems. However, characterizing altermagnets through charge responses such as the AHE or magneto-optical effects encounters a fundamental limitation: these responses vanish in the absence of SOC. More generally, since both the Hall vector and net magnetization transform as time-odd pseudovectors, their effects often appear together. This necessitates distinguishing between altermagnetic and weak-ferromagnetic contributions~\cite{Hariki2025}.

X-ray magnetic circular dichroism (XMCD), which probes transitions from atomic-like core levels to valence states, 
offers an advantage in this respect. Although SOC is still required for XMCD, it is typically dominated by the strong SOC 
of the core orbitals, while valence SOC plays only a minor role. Moreover, core–valence exchange is irrelevant when the core levels 
are filled. Assuming this exchange can be neglected in the final state containing a core hole, the valence shell effectively 
regains spin rotational symmetry. Together with the well justified assumption of rotational symmetry in the core Hamiltonian, these conditions 
restore the applicability of spin-group classification and define what we refer to as the free valence spin (FVS) approximation.

The XMCD spectra for light propagating along the direction $\hbk$ is obtained as a projection $2\bh(\omega)\cdot\hbk$ of
the frequency dependent Hall vector ${{\bf h}(\omega) = \Im(\sigma^a_{zy}(\omega), \sigma^a_{xz}(\omega), \sigma^a_{yx}(\omega))}$.
The Hall vector ${\bf h}(\omega)$ depends on the orientation of the magnetic moments in the sample. In antiferromagnets and altermagnets
this orientation is described by the N\'eel vector $\bL=\bbm_A-\bbm_B$, defined as the difference of sublattice magnetizations.

Previously, we studied XMCD in MnTe~\cite{Hariki2024b} and rutiles RuO$_2$~\cite{Hariki2024a}, MnF$_2$~\cite{MnF2} and NiF$_2$~\cite{Hariki2025}. 
XMCD is present and well described
by theory in the experimental geometry, $\bL\parallel \langle 1\bar{1}00\rangle$ and $\bk\parallel [0001]$, of MnTe~\cite{Hariki2024b}. Moreover,
the weak-ferromagnetic contribution was shown to be negligible. Interestingly, XMCD vanishes identically in the FVS approximation.
In the rutile structure, XMCD is allowed for $\bL$ in the $ab$-plane, a geometry realized in NiF$_2$. We have shown that
XMCD is present in the FS approximation and a peculiar mirror image relationship exists between the Hall vector $\bh(\omega)$ and the N\'eel
vector $\bL$. Numerical simulations~\cite{Hariki2025} revealed that this relationship preserved, with a fraction of per cent accuracy, even for the
full Hamiltonian.

In this Letter we present the general derivation of the relationship between the Hall and N\'eel vectors in the FVS approximation using
group theoretical methods. We recover the previous results for specific geometries in MnTe and rutile altermagnets as special cases of 
simple general rules.

The key observation is a linear relationship between the Hall vector $\bh(\omega)$ and the N\'eel vector $\bL$. 
\begin{widetext}
\begin{equation}
\label{eq:main}
    \begin{split}
        h_A^\gamma(\omega)&=\sum_{i,f}\epsilon_{\alpha\beta\gamma}\langle\Psi_i|\oT^\alpha_A|\Psi_f\rangle
        \langle\Psi_f|\oT^\beta_A|\Psi_i\rangle\delta(\omega-E_f+E_i)\\
        &=\sum_{i,f}\epsilon_{\alpha\beta\gamma}(T^\alpha_{m\mu})^*T^\beta_{m'\mu'}
         \langle\Psi_i|p^\dagger_{\mu s}d^{\phantom\dagger}_{ms}|\Psi_f\rangle
        \langle\Psi_f|d^\dagger_{m's'}p^{\phantom\dagger}_{\mu' s'}|\Psi_i\rangle\delta(\omega-E_f+E_i) \\
        &=\sum_{i,f}\sum_{j,j_z}\epsilon_{\alpha\beta\gamma}(T^\alpha_{m\mu})^*T^\beta_{m'\mu'}
         \langle\psi_i|d^{\phantom\dagger}_{ms}|\psi_f\rangle
        \langle\psi_f|d^\dagger_{m's'}|\psi_i\rangle
        \langle\emptyset|p^\dagger_{\mu s}|j,j_z\rangle
        \langle j,j_z|p^{\phantom\dagger}_{\mu' s'}|\emptyset\rangle\delta(\omega-\tilde{E}_f-E^c_j+U_{pd}+E_i)\\
        &=\sum_j \mD_{ms,m's'}(\omega+\Delta_j)\Lambda^\gamma_{ms,m's'}(j)
        =\sum_j \mD^t_{m,m'}(\omega+\Delta_j)\Lambda^\gamma_{ms,m's'}(j)\sigma^{\gamma'}_{ss'}\hat{m}_A^{\gamma'}\\
        &\equiv\Omega^{\gamma\gamma'}_A(\omega)\hat{m}_A^{\gamma'}.
    \end{split}
\end{equation}
\end{widetext}
Here $\oT^\alpha$ is $\alpha$'s cartesian component of the dipole operator on lattice site $A$, $\epsilon_{\alpha\beta\gamma}$ is the 
Levi-Civita symbol, $|\Psi_i\rangle$ and $|\Psi_f\rangle$ are the initial eigenstate and final eigenstates with core hole on site $A$, respectively, of the absorption
process, and $E_f$ and $E_i$ are the corresponding energies. Summation over repeated indices is implied. 
On the second line we introduce the creation and anihilation operator for the valence states $d_{ms}$ and core states $p_{\mu s}$ with the orbital indices $m$ and $\mu$, and spin index $s$ (we do not show the site index $A$ for sake of simplicity). Assuming that the core-valence interaction has only the monopole term
the wave functions factorize into the valence part $|\psi\rangle$ and the core part. The latter is full shell state $|\emptyset\rangle$
in the initial states and a core hole $|j,j_z\rangle$ labeled by the total angular momentum and its projection. The energy
of the final state is a sum of the valence energy $\tilde{E}_f$ and the core-hole energy $E^c_j$, which does not depend on $j_z$~\footnote{By the rotation symmetry of the core Hamiltonian.},
and a core-hole interaction $U_{pd}$, which is the same for all final states. The sums over $j_z$, $\mu$, $\mu'$ and cartesian indices $\alpha$ and $\beta$
leads to a universal matrix $\Lambda^\gamma_{ms,m's'}(j)$ for each absorption edge labeled by $j$. 
The sum over the initial and final states yields the material specific $\omega$-displaced spectral function for the unoccupied valence states
$\mD_{ms,m's'}(\omega+\Delta_j)$ (site index is not shown). The Hall vector is the trace over the product of these matrices. In the absence of
valence SOC $\mD_{ms,m's'}(\omega)$ has a form tensor product of spin and orbital parts 
\begin{equation}
        \bmD_A^{\phantom{s}}(\omega)=\bmD_A^s(\omega)\otimes I + \bmD_A^t(\omega)\otimes (\hat{\mathbf{m}}_A\cdot\boldsymbol{\sigma}),
\end{equation}
where $\bmD_A^s(\omega)$ and $\bmD_A^t(\omega)$ matrices in the orbital space, while $I$ and $\hat{\mathbf{m}_A}\cdot\boldsymbol{\sigma}$ are
$2\times 2$ spin matrices: $I$ is the identity matrix, $\sigma^{\gamma}$ are the Pauli matrix and $\hat{\mathbf{m}}$ 
is the unit vector in the direction of the local spin moment. The contribution of the first term to (\ref{eq:main}) vanishes
assuming that time-reversal symmetry is broken only in the spin space. Substituting the second term in (\ref{eq:main}) 
we find the linear relationship between $\bh_A(\omega)$ and $\hat{\mathbf{m}}_A$.
The total $\bh(\omega)$ is obtained by summation over all atoms.

The question to address is which form of $\boldsymbol{\Omega}_A(\omega)$ is permitted by the symmetry of the system.  We first analyze the contribution from a single magnetic sublattice and then sum over the sublattices with opposite $\hat{\mathbf{m}}$. 
Up to this point, we have relied on the fact that 
the Hamiltonian in FVS approximation is invariant under arbitrary rotation 
of 
the valence spin can be rotated independently of both the core spin and the lattice, which allowed us to demonstrate a linear relationship between $\mathbf{h}_A(\omega)$ and $\hat{\mathbf{m}}_A$.

In the following we consider the diagonal subgroup of $H_s$ of $H\otimes SU(2)$, formed by orbital transformations $H$ and spin rotations $SU(2)$, i.e. we pair the same rotations/reflections in the orbital and spin space. $H$ is the subgroup of the crystallographic point group, which preserves the magnetic sublattice structure. Moreover, the matrices $\bmD_A(\omega)$ and $\boldsymbol{\Lambda}(j)$ couple only harmonics of the same angular momenta and thus are insensitive 
to spatial inversion. As a result, they do not distinguish point groups corresponding to the same Laue group and $H$ can be taken as the halving group of the spin Laue group as defined in Ref.~\onlinecite{Smejkal22}. In systems with one atomic site per magnetic sublattice 
$H$ is the site symmetry respecting the magnetic order. If there are multiple 
atoms per magnetic sublattice we extend the summation of over the final states
to include core-hole excitation to all atoms on a sublattice wit the same $\bbm$, i.e., $\bh_A(\omega)$ becomes the sublattice sum and 
$A$ is understood as the sublattice index in the rest of the paper. 
The transformation of $\bh_A(\omega)$ under $U\in H_s$ reads
~\footnote{Here we need the simultaneous orbital/spin rotation in order to express an operation $U$ acting on $m$ and $\mu$ indices of $T^\alpha_{m\mu}$
as an operation $D(U)$ acting on $\alpha$.}
%\begin{widetext}
\begin{equation}
    \begin{split}
        &h_A^\gamma(U)
        \equiv\sum_{i,f}\epsilon_{\alpha\beta\gamma}\langle U\Psi_i|\oT^\alpha|U\Psi_f\rangle
        \langle U\Psi_f|\oT^\beta|U\Psi_i\rangle
        %\delta(\omega-E_f+E_i)
        \delta_{fi}\\
        &=\sum_{i,f}\epsilon_{\alpha\beta\gamma}\langle \Psi_i|U^{-1}\oT^\alpha U|\Psi_f\rangle
        \langle \Psi_f|U^{-1}\oT^\beta U|\Psi_i\rangle
        \delta_{fi}%\delta(\omega-E_f+E_i)
        \\    
        &=D_{\xi\gamma}^{(R)}(U)h_A^{\xi}.
    \end{split}
\end{equation}
Here we have used that $|\Psi\rangle$ and $|U\Psi\rangle$ are eigenvectors with the same eigenvalue and abbreviate the delta fuction
of Eq.~\ref{eq:main} with $\delta_{fi}$. Since the dipole operators $\oT^\xi$ transform as vectors their cross product transforms as pseudovector
representation $R$. The spin moment $\bbm_A$ transforms as a pseudovector thanks to the algebra of Pauli matrices
\begin{equation}
    \begin{split}
        m_A^{\gamma}(U)
        &\equiv\sum_{i}\langle U\Psi_i|d^\dagger_{m\alpha}d^{\phantom\dagger}_{m\beta}|U\Psi_i\rangle\sigma^\gamma_{\alpha\beta}\\    
       &=D^{(R)}_{\xi\gamma}(U)m_A^{\xi}.
    \end{split}
\end{equation}
Since both $\bh_A$ and $\hat{\bbm}_A$ transform as the pseudovector representation $R$ of $H$ we have to find the invariants of
the tensor product $R^*\otimes R$ in order to uncover the form of $\boldsymbol\Omega_A(\omega)$ of Eq.~\ref{eq:main}. Once this is done we can obtain total Hall vector as a sum over sublattices
$\bh(\omega)=\bh_A(\omega)+\bh_B(\omega)$. The contribution $\boldsymbol{\Omega}_B(\omega)$ of sublattice $B$  is obtained transforming $\boldsymbol{\Omega}_A(\omega)$ by an operation $\mathbf{g}$, which maps sublattice $A$ to sublattice $B$ and reversing the sign, which reflects opposite orientation of the spin moment $\bbm_B=-\bbm_A$. The final expression for the relationship between the Hall vector $\bh(\omega)$ and the N\'eel vector 
reads
\begin{equation}
\label{eq:total_h}
        \bh(\omega)=\frac{1}{2}\left(\boldsymbol{\Omega}(\omega)-
        \mathbf{g}\boldsymbol{\Omega}(\omega)\mathbf{g}^T\right) \hat{\bL}.
\end{equation}
We have dropped the sublattice index; $\boldsymbol{\Omega}(\omega)$ is a selected sublattice contribution and $\mathbf{g}$ is the mapping to the sublattice with opposite spin.
The choice of $\mathbf{g}$ is not unique, but the above result is by the coset property of $gH$.
In antiferromangets with sublattices connected by translation or inversion the operation $\mathbf{g}$ reduces
to identity matrix $\pm I$ and thus $\bh(\omega)$ of (\ref{eq:total_h}) vanishes. However, as we show below (\ref{eq:total_h})
may lead to vanishing $\bh(\omega)$ even if it is allowed by magnetic point group. This is a consequence of higher symmetry of the
FVS approximation. In such case the inclusion of core-valence multiplets and possibly valence SOC is necessary  to get finite XMCD.

\subsection{Rutile structure}
Materials with rutile strucutre such as $RuO_2$ or transition metal difluorides are much studied altermagnets. With 
the crystallographic point group $D_{4h}$ the magnetic order fits into the two atomic unit cell leading to
the halving group is $H=D_{2h}$. The pseudovector representation splits into three one-dimensional 
irreducible reprepresentations $R=B_{1g}\oplus B_{2g}\oplus B_{3g}$ with basis functions $R_x$, $R_y$ and $R_z$, for the
coordinate axes aligned with the $C_2$ axes. This leads to three invariants ($A_{1g}$ representations) in $R^*\otimes R$
and the single site $\boldsymbol{\Omega}(\omega)$ of the form
\begin{equation*}
    \boldsymbol{\Omega}(\omega) = \begin{pmatrix} a(\omega) & 0 & 0 \\ 0 & b(\omega) & 0 \\0 & 0 &c(\omega) \end{pmatrix}, \quad
    \mathbf{g} = \begin{pmatrix} 0 & 1 & 0 \\ -1 & 0 & 0 \\0 & 0 & 1 \end{pmatrix}.
\end{equation*}
The mapping between the sublattices is facilitated by a rotation of $90\deg$ about the $c$ -axis.
Evaluating (\ref{eq:total_h}) we arive at the result in Tab.~\ref{tab:symmetry-data}.
Finally, transformation to the common coordinate setting with $C_2\parallel \langle 1 1 0\rangle$ we obtain 
\begin{equation*}
    \bh(\omega) = \begin{pmatrix} 0 & \tilde{a}(\omega) & 0 \\ \tilde{a}(\omega) & 0 & 0 \\0 & 0 & 0 \end{pmatrix}\hat{\bL}, \quad
    \tilde{a}(\omega) = a(\omega)-b(\omega).
\end{equation*}
This shows that there is a single irreducible spectral function and a mirror relationship between the Hall and N\'eel vectors derived in Ref.~\onlinecite{MnF2}. Interestingly, when restricted to the $ab$-plane the relationship is invertible and thus the Hall vector determines the N\'eel vector uniquely. For example, measuring the polar angle of maximizing XMCD in NiF$_2$ at a selected energy can be used
to distinguish the four $\langle 100\rangle$ domains.

As shown in Refs.~\onlinecite{MnF2,Hariki2025} for MnF$_2$ and NiF$_2$ the inclusion of core-valence interaction and valence SOC
modifies the shape of the XMCD spectra. However, the above relationship between the Hall and N\'eel vectors remains fullfilled
to fraction of a per cent accuracy. The main effect of valence SOC in NiF$_2$ is canting of local moments leading to
a small net magnetization. The total XMCD signal in NiF$_2$ can be very accurately described as a sum of the above contribution 
proportional to $\bL$ and the ferromagnetic contribution 
%$\bh_M(\omega)=A_M(\omega)\mathbf{M}$ 
proportional to the net magnetization %$\mathbf{M}$
~\cite{Hariki2025}.

\subsection{NiAs structure}
NiAs is another common structural type among popular altermagnets with representatives such as MnTe or CrSb.
Similar to the rutile structure the magnetic order in these materials fits to the two atomic crystallographic 
unit cell with the point group $D_{6h}$ leading to the halving group $H=D_{3d}$. The pseudovector representation splits into two  
irreducible reprepresentations $R=A_{2g}\oplus E_{g}$ with basis functions $R_z$ and $(R_x, R_y)$. This leads to two invariants
in $R^*\otimes R$ representation and 
\begin{equation*}
   \boldsymbol{\Omega}(\omega)=\begin{pmatrix} a(\omega) & 0 & 0 \\ 0 & a(\omega) & 0 \\0 & 0 &c(\omega) \end{pmatrix},\quad
%    \mathbf{A} =\begin{pmatrix} \cos(\pi/3) & \sin(\pi/3) & 0 \\ -\sin(\pi/3) & \cos(\pi/3) & 0 \\0 & 0 & 1 \end{pmatrix}.
    \mathbf{g} =\begin{pmatrix} \tfrac{1}{2} & -\tfrac{\sqrt{3}}{2} & 0 \\ 
    \tfrac{\sqrt{3}}{2} & \tfrac{1}{2} & 0 \\0 & 0 & 1 \end{pmatrix}.
\end{equation*}
Here, we have chosen the $C_{6z}$ operation for $A$. It leaves $\boldsymbol{\Omega}(\omega)$ unchanged and thus $\bh(\omega)=0$ upon
the summation over the magnetic sublattices. 

In MnTe with $\bL||\langle 1\bar{1}00\rangle$ a finite XMCD exists~\cite{Hariki2024b, Yamamoto} with $\bh(\omega)\parallel c$. Nevertheless, as shown in Ref.~\onlinecite{Hariki2024b} both analytically and numerically the core-valence exchange is crucial~\footnote{Valence SOC alone leads to
a finite XMCD signal, but its contribution is order of magnitude smaller.} In CrSb with  $\bL||\langle 0001\rangle$ $\bh(\omega)=0$ even
with full Hamiltonian as dictated by the magnetic point group.

\begin{table*}[ht]
\centering
\caption{The irreducible components of the x-ray Hall vector according to Eq.~\ref{eq:total_h} for a single sublattice,
$\boldsymbol{\Omega}$, and for the entire system $\tfrac{1}{2}(\boldsymbol{\Omega}-
 \mathbf{g}\boldsymbol{\Omega}\mathbf{g}^T)$. The Laue group $G$, the halving
        subgroup $H$ and the symmetry operation $\mathbf{g}$ connecting the two sublattices as well as the candidate materials follow Fig.~2 of Ref.~\cite{Smejkal22}} .
\label{tab:symmetry-data}
%\resizebox{\textwidth}{!}{  % scales table if too wide
\renewcommand{\arraystretch}{1.1}
\begin{tabular}{l | l| l| c  c |l|l }
\toprule
  $G$ & $H$ & $\mathbf{g}$ & $\boldsymbol{\Omega}$ &  $\tfrac{1}{2}(\boldsymbol{\Omega}-
        \mathbf{g}\boldsymbol{\Omega}\mathbf{g}^T)$\\
\hline
  $D_{2h}$ & $C_{2h}$ & $C_{2x}$ &  
  $\begin{pmatrix} a & b & 0 \\ c & d & 0 \\0 & 0 & e \end{pmatrix}$ &
  $\begin{pmatrix} 0 & b & 0 \\ c & 0 & 0 \\0 & 0 & 0 \end{pmatrix}$ & 
  & \parbox{2.5cm}{FeSb$_2$, La$_2$CuO$_4$}\\

  $C_{4h}$ & $C_{2h}$ & $C_{4z}$ &  
  $\begin{pmatrix} a & b & 0 \\ c & d & 0 \\0 & 0 & e \end{pmatrix}$ &
  $\begin{pmatrix} \tilde{a} & \tilde{b} & 0 \\ \tilde{b} & -\tilde{a} & 0 \\0 & 0 & 0 \end{pmatrix}$ & 
  \parbox{2cm}{$\tilde{a} = \tfrac{a-d}{2}$\\ $\tilde{b} = \tfrac{b+c}{2}$}& \\

 $D_{4h}$  & $D_{2h}$ & $C_{4z}$ & 
  $\begin{pmatrix} a & 0 & 0 \\ 0 & b & 0 \\0 & 0 & c \end{pmatrix}$ &
  $\begin{pmatrix} \tilde{a} & 0 & 0 \\ 0 & -\tilde{a} & 0 \\0 & 0 & 0 \end{pmatrix}$ &
  \parbox{2cm}{$\tilde{a} = \tfrac{a-b}{2}$ } &
  \parbox{2cm}{RuO$_2$, NiF$_2$, \\ MnF$_2$}
  \\

  \parbox{0.5cm}{$D_{4h}$ \\ $D_{6h}$} & \parbox{0.5cm}{$C_{4h}$ \\ $C_{6h}$} &  \parbox{0.5cm}{$C_{2x}$\\$C_{21}$} & 
  $\begin{pmatrix} a & 0 & 0 \\ 0 & a & 0 \\0 & 0 & b \end{pmatrix}$ &
  0 & &  \parbox{2cm}{KMnF$_3$}
  \\
  $C_{2h}$ & $C_i$ & $C_{2z}$ &
  $\begin{pmatrix} a & b & c \\ d & e & f \\ g & h & i \end{pmatrix}$ &
  $\begin{pmatrix} 0 & 0 & c \\ 0 & 0 & f \\ g & h & 0 \end{pmatrix}$ & &
   \parbox{2cm}{CuF$_2$}
  \\

 \parbox{0.5cm}{$D_{3d}$ \\ $D_{6h}$}  & \parbox{0.5cm}{$S_{6}$ \\ $D_{3d}$} &  \parbox{1.5cm}{$C_{21}$, $C_{6z}$ \\$C_{21}$} & 
  $\begin{pmatrix} a & 0 & 0 \\ 0 & a & 0 \\0 & 0 & b \end{pmatrix}$ &
  0 & & \parbox{2cm}{ CoF$_3$, Fe$_2$O$_3$\\CrSb, MnTe} \\
  $O_h$ & $T_{d}$ & $C_{4z}$ &
  $aI
  %\begin{pmatrix} a & 0 & 0 \\ 0 & a & 0 \\0 & 0 & a \end{pmatrix}
  $ &
  $0$ &
  &
\end{tabular}
\end{table*}

Using group-theoretical arguments, we have derived a relationship between the x-ray Hall vector and the Néel vector within the FVS approximation. Our approach generalizes and streamlines earlier derivations for specific cases such as MnTe~\cite{Hariki2024b} and MnF$_2$\cite{MnF2}, which relied on elementary linear algebra. While our focus is on altermagnets, the results also apply to ferromagnets—where the second sublattice is absent—and can be readily extended to non-collinear magnets with multiple magnetic sublattices\cite{Wimmer19, Mn3Sn}. We applied this method to representative altermagnetic structures. For some, such as rutile, the FVS approximation yields a non-zero Hall vector, while for others, such as NiAs, the Hall vector vanishes identically, requiring symmetry lowering via core-valence exchange or valence SOC to make a Hall response admissible. Previous simulations on MnF$_2$ and NiF$_2$\cite{Hariki2025} showed that although the FVS approximation does not accurately reproduce the XMCD spectral shape, the symmetry relations summarized in Table\ref{tab:symmetry-data}—derived from the FVS framework—are satisfied to high precision. The FVS-based analysis thus serves as a valuable starting point for understanding the link between XMCD and magnetic moment orientation in magnetic materials.

\bibliography{main} 

\begin{acknowledgements}
We thank Atsushi Hariki, Mizuki Furo and Anna Kauch for numerous discussions and critical reading of the manuscript.
This work was supported by the project Quantum materials for applications in sustainable technologies (QM4ST), funded as project No. CZ.02.01.01/00/22\_008/0004572 by Programme Johannes Amos Commenius, call Excellent Research and
by the Ministry of Education, Youth and Sports of the Czech Republic through the e-INFRA CZ (ID:90254). 
\end{acknowledgements}

%\bibliography{main} 

\end{document}